\documentclass[prb,aps,twocolumn,amsmath,amssymb,showpacs,superscriptaddress]{revtex4}

\usepackage{dcolumn}
\usepackage{bm}
\usepackage{graphicx}
\usepackage{psfrag}
\usepackage{subfigure}

\newcommand{\cdag}{c^\dagger}
\newcommand{\cnod}{c^{\phantom{\dagger}}}

\begin{document}
 \title{Magnetism of one-dimensional Wigner lattices and its impact on charge order}
 \author {M. Daghofer}
 \affiliation{ Max-Planck-Institut f\"ur Festk\"orperforschung,
               Heisenbergstrasse 1, D-70569 Stuttgart, Germany }
\affiliation{ Materials Science and Technology Division, Oak
  Ridge National Laboratory, Oak Ridge, Tennessee 37831, USA and
  Department of Physics and Astronomy, The University of Tennessee,
  Knoxville, Tennessee 37996, USA}
\email{M.Daghofer@fkf.mpg.de}
\author {R. M. Noack}
 \affiliation{ Philipps Universit\"at Marburg, D-35032 Marburg, Germany } 
\author {P. Horsch}
 \affiliation{ Max-Planck-Institut f\"ur Festk\"orperforschung,
              Heisenbergstrasse 1, D-70569 Stuttgart, Germany }
 \date{\today}

\begin{abstract}

The magnetic phase diagram of the quarter-filled generalized Wigner lattice
with nearest- and
next-nearest-neighbor hopping $t_1$ and $t_2$ is explored.
We find a region
at negative  $t_2$ with fully saturated ferromagnetic ground states
that we attribute to kinetic exchange. Such interaction 
disfavors antiferromagnetism at $t_2 <0$  and stems from
virtual excitations across the charge gap of the Wigner lattice, which is
much smaller than the Mott-Hubbard gap $\propto U$.
Remarkably, we find a strong dependence of the charge
structure factor on magnetism even in the limit $U\rightarrow\infty$, 
in contrast to the expectation that charge ordering
in the Wigner lattice 
regime
should be  well described by spinless fermions. 
Our results, obtained using the density-matrix renormalization group
and exact diagonalization, can be transparently explained by means of an effective
low-energy Hamiltonian.
\end{abstract}


\maketitle
\section{Introduction}\label{sec:intro}

In a Wigner lattice (WL), long-range 
Coulomb repulsion dominates over the kinetic energy of electrons
and leads to strong and well-defined charge 
order.~\cite{Wigner34} 
Originally
introduced for the electron gas with a homogeneous neutralizing background,
this concept
was generalized to  electrons on a lattice by Hubbard.~\cite{Hubbard78}
Evidence for quasi one-dimensional (1D) Wigner lattices has been found in 
organic~\cite{Hir98,Nad06,Cla07,kakiuchi:066402} 
and anorganic~\cite{Cox98,Mat99,Kud05,Horsch05} 
chain compounds, 
nanowires~\cite{Rah07}  and in
carbon nanotubes.\cite{Des08}
Low-dimensional WLs are favored by reduced screening.\cite{Hubbard78,Capponi00}
Moreover, strong correlations induced by large local Hubbard interaction $U$
suppress screening further and protect the long-range nature of
the Coulomb repulsion.\cite{Horsch05}
It is, however, not straightforward to
distinguish a true WL from a quantum-mechanical charge-density wave (CDW)
simply on the basis of the periodicity of the charge modulation.
In fact, it turns out that the modulation period of the WL coincides
with that of the $4k_F$ CDW.~\cite{Hubbard78,Hir83,Sch93,Nad06,schuster:045124} 
This may be surprising, as the microscopic origin of the charge order
in the two cases is fundamentally different:\cite{kakiuchi:066402}
(i) The mechanism for the WL is based solely on the classical Coulomb
repulsion and is dependent only on the charge of electrons and not on their
fermionic nature. The periodicity follows simply from the configuration
of charges with minimal energy.
(ii) Instead, the quantum
mechanical CDW depends on the Fermi surface topology and the
instability and modulation reflects the Fermi momentum $k_F$.
 
If we allow for nearest--neighbor (NN) and next--nearest--neighbor
(NNN) hoppings $t_1$ and $t_2$,
we arrive at an even more interesting model which, depending on the relative 
size and sign of  $t_1$ and $t_2$, may have an electron dispersion with
two minima instead of one.\cite{Fab96,Ari98}
Actually,  it has been proposed that the edge-sharing CuO chain compounds
are described by such models with $|t_2|>|t_1|$.~\cite{Horsch05}
It is then immediately obvious that in the case of a four-Fermi-point topology
the periodicities of WL and CDW no longer coincide.
We note that the experimentally observed charge modulations in the edge-sharing compounds
Na$_{1+x}$CuO$_2$~\cite{Horsch05,Sofin05,Sma07,Rai08} 
%
are strong and their periodicity  consistent only with that of the WL. 
Another possible type of instability in the presence of strong correlations, 
namely, the 2$k_F$ Peierls and spin-Peierls modulations, 
\cite{Mal03,schuster:045124}
which requires a distortion of the lattice, appears to be ruled out
in these systems.
Doped edge-sharing chains are also building blocks
of the Ca$_{2+x}$Y$_{2-x}$Cu$_{5}$O$_{10}$\cite{Kud05,Mat05} 
and the  Sr$_{14-x}$Ca$_{x}$Cu$_{24}$O$_{41}$\cite{Cox98,Mat99,Iso00}
systems, and pronounced charge order has been observed in these compounds as well.

Here we shall investigate the intrinsic mechanisms for the magnetism 
of generalized Wigner lattices. 
It is well known that NNN hopping $t_2$ has nontrivial consequences for
magnetism in  the 1D Hubbard model at general filling and may lead to
ferromagnetic (FM) states in certain cases.
\cite{Mue95,Pen96,Pie96, PhysRevB.58.2635,Nis08}
It should be kept in mind that, according to the Lieb-Mattis theorem, \cite{PhysRev.125.164}
ferromagnetism is excluded at any filling in the 1D Hubbard model
with NN hopping (i.e., $t_2=0$).
For the 1D Hubbard model with NN and NNN
hopping, Pieri {\em et al.}\ \cite{Pie96} and 
Daul and Noack \cite{PhysRevB.58.2635} found that 
ferromagnetic ground states appeared above a critical $U$ in 
those regions of the $t_1$-$t_2$ plane where four Fermi points exist. 
These results were obtained in the metallic regime where the relevance
of Fermi surface topology is suggestive;
however, the implications for the
localized electrons of a Wigner crystal are unclear.

The magnetism of generalized WLs is typically discussed in terms of effective 
Heisenberg models
where the position of the spins is dictated by the charge order pattern of the WL.
\cite{Horsch05,Sch05,Kli06}
The prevailing superexchange interactions are antiferromagnetic. 
However, there are also ferromagnetic couplings  in edge-sharing
chains due to the Hund interaction at the oxygen ligands that may
be larger than the 
AF interactions and render, e.g., the nearest-neighbor interaction
$J_1$ ferromagnetic. \cite{Miz98,Tor99,Horsch05}
These features lead to frustration, and the resulting helical spin states have been 
observed in the spin-1/2 edge-sharing
chain compounds LiCu$_2$O$_2$\cite{Mas04} and
NaCu$_2$O$_2$.\cite{Cap05,Dre06,Dre07} 

In this paper, we show that, for a Wigner crystal at quarter-filling,
there is another  intrinsic mechanism that may
lead to ferromagnetism. 
By means of a density-matrix renormalization group (DMRG) study 
of the $t_1-t_2$ Hubbard-Wigner model,
which includes local Hubbard $U$ and long-range 
Coulomb interactions $V_l=V/l$, we show that there is a regime of
fully polarized FM states at negative $t_2$. 
Subsequently, we derive an effective magnetic Hamiltonian
for the Wigner-lattice regime, i.e. $|t_1|,|t_2| \ll V <U$, and show that the
emergence of ferromagnetism can be explained by an effective
\emph{kinetic exchange} mechanism mediated by NNN hopping
$t_2$. The associated magnetic exchange constant $\propto t_1^2
t_2/\epsilon_0^2$ depends on the sign of $t_2$ and therefore kinetic
exchange is found to favor ferromagnetism for negative $t_2$ and
antiferromagnetism for positive $t_2$. Kinetic exchange
involves excitations across the charge gap $\epsilon_0$ of 
the WL but not across the usually much larger Mott-Hubbard gap $\sim
U$, as is the case for AF superexchange or for many realizations of FM
three-particle ring exchange.~\cite{0370-1328-86-5-301} 
The charge gap $\epsilon_0$ of the generalized WL depends sensitively on
the commensurability with the underlying crystalline lattice.
At quarter-filling this gap is particularly large $\epsilon_0\sim V/2$.

If the charge gap $\epsilon_0\sim V/2 $ of the WL is much larger than 
the hopping amplitudes  $t_1$ and $t_2$ and any expected
magnetic couplings, a separation of charge and magnetic energy scales appears
straightforward. Hence, charge ordering in a WL is usually discussed in
terms of spinless fermions. Magnetism, e.g., antiferromagnetic (AF)
superexchange or  ferromagnetic (FM) Hund's rule~\cite{Horsch05} and
three-site ring exchange,~\cite{klironomos:075302} is then treated as
a perturbation given a particular charge-ordering pattern. One would, however,
not expect the magnetic order to have a strong impact on the
underlying charge order, because the magnetic energy scale is so much
smaller than the dominant Coulomb repulsion for all these processes.~\cite{Valenzuela03} 
The motivation for this work was the initial observation that the
charge structure of the WL, measured by the charge structure factor
$N(q)$ at $q=\pi$, is strongly affected by electron spin, in
disagreement with the calculation for spinless fermions.
Yet there is a region in the $t_1-t_2$ phase diagram at negative $t_2$
where $N(\pi)$ is the same 
for spinless fermions and for fermions with spin, and, moreover,
$N(\pi)$ does not depend on $t_2$ in that parameter range.
The obvious conjecture is that the ground state should be
fully spin-polarized in this regime. 

We show here that, due to the kinetic exchange mechanism, these
FM ground states emerge and that the kinetic exchange
processes have  a surprisingly strong impact on the charge
ordering
in spite of the classical origin of the WL. Indeed, the AF
state  at $t_2 > 0$ has dramatically weaker  charge order than the
ferromagnetic or spinless states. 
While the charge order is reduced  $\propto t_2$ for positive $t_2$,
it does not depend on $t_2$ in the FM regime  
$t_2^a >t_2 > t_2^b$ with negative $t_2$ . 
Remarkably, for 
negative $t_2$ values below $ t_2^b$, AF reappears,  yet
the charge order then increases with increasing modulus  $|t_2|$. 
The boundaries of
the FM phase follow from the effective spin Hamiltonian as
 $t_2^a \sim -3 t_1^2/U $ and $t_2^b \sim -(U/\epsilon_0^2)
t_1^2 $, and match the magnetic phase boundaries found using the 
DMRG.
This peculiar behavior is due to a purely quantum
effect involving destructive interference 
of kinetic exchange processes in the FM state due to
the Pauli principle and constructive interference for the AF case.

The paper is organized as follows:
After introducing the Hubbard-Wigner 
Hamiltonian in Sec.~\ref{sec:model}, we present
results for the charge structure factor for spinless fermions
interacting via long-range Coulomb interaction in Sec.~\ref{sec:charge}.
We shall see that the results for spinless fermions
coincide with results for electrons with spin
in some region of the $t_1$-$t_2$ phase diagram,
while they are substantially different in other parts of the
phase diagram. We derive an effective Heisenberg Hamiltonian in
Sec.~\ref{sec:spin} and show that the magnetic phase diagram,
i.e., the appearance of the fully saturated FM phase
and its phase boundaries, can be naturally explained. Next we show
that the peculiarities found numerically for the the charge structure
factor find a straightforward analytical description in the
framework of the effective Hamiltonian. 
Finally, we discuss and summarize our results in Sec.~\ref{sec:conclusions}.
\begin{figure}
  \centering
{ \psfrag{a}{\hspace*{-1em}(a)}
  \psfrag{b}{\hspace*{-1em}(b)}
  \psfrag{c}{\hspace*{-1em}(c)}
  \psfrag{d}{\hspace*{-1em}(d)}
  \psfrag{t1}{$-t_1$}
  \psfrag{t2}{$-t_2$}
  \includegraphics[width=0.17\textwidth]{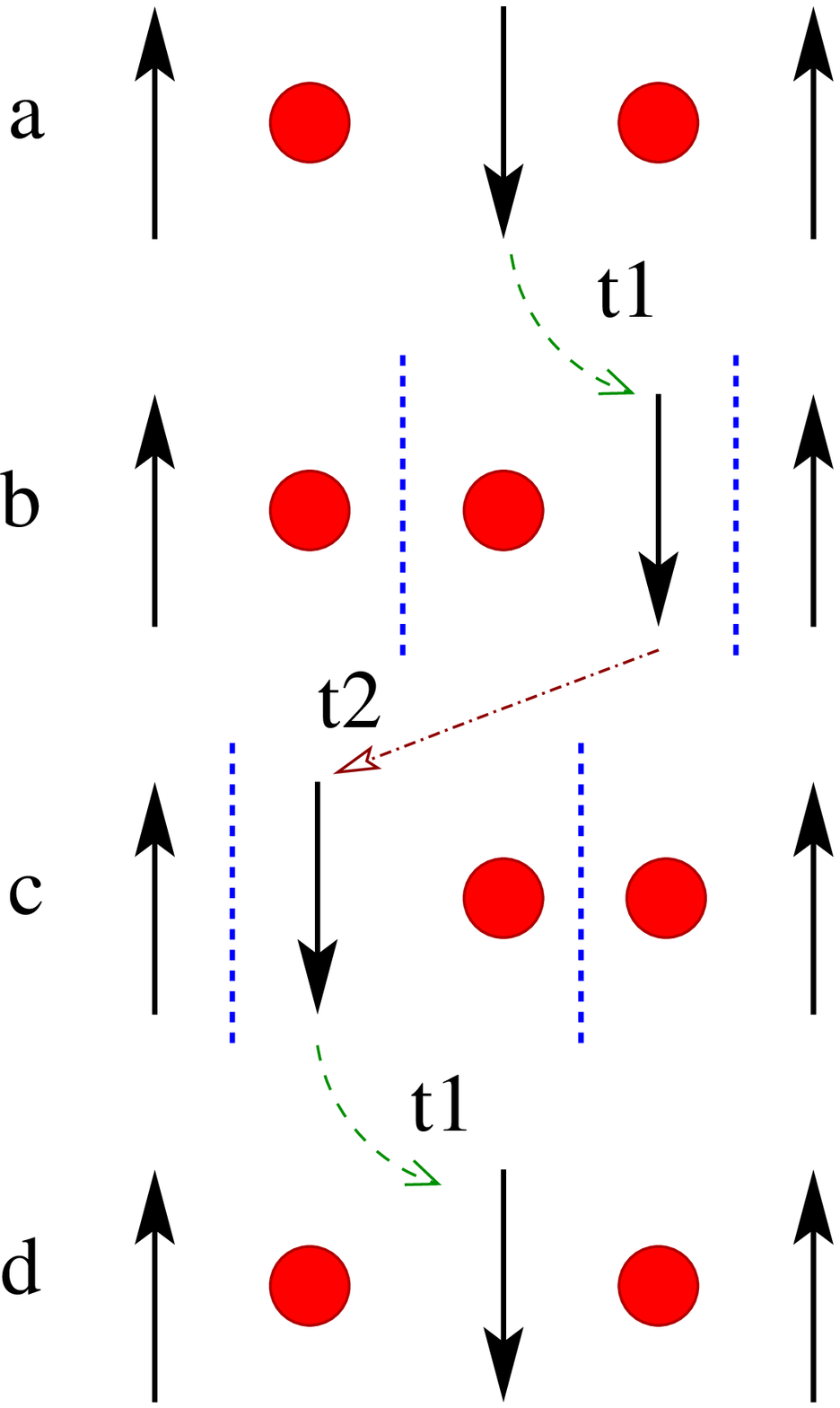}}\hspace*{3.5em}
{ \psfrag{a}{\hspace*{-1em}(e)}
  \psfrag{b}{\hspace*{-1em}(f)}
  \psfrag{c}{\hspace*{-1em}(g)}
  \psfrag{d}{\hspace*{-1em}(h)}
  \psfrag{t1}{$-t_1$}
  \psfrag{t2}{$+t_2$}
  \includegraphics[width=0.17\textwidth]{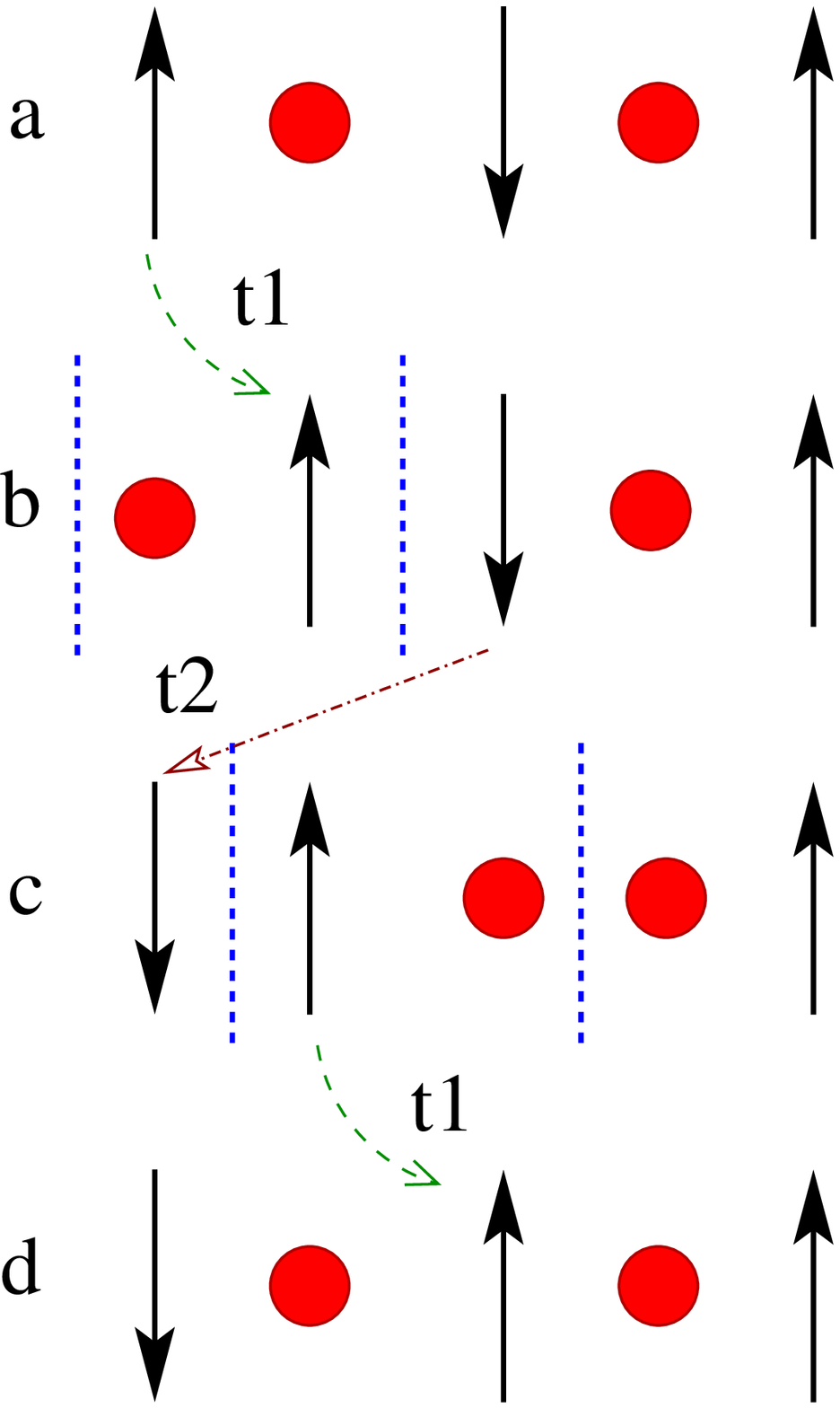}}\\
\caption{
  (Color online) Schematic depiction of relevant  
  $\propto t_1^2 t_2/\epsilon_0^2$ kinetic exchange
  processes. We start from perfect charge order (a and e), where
  circles denote empty sites (Cu$^{3+}$ atoms in
  Na$_{1+x}$CuO$_2$), and arrows denote occupied sites (Cu$^{2+}$). NN
  hopping $t_1$
  induces excitations (a$\to$b and e$\to$f) with  
  two domain walls (dashed lines) and cost $\epsilon_0$. 
  Two different $t_2$ processes, one with (f$\to$g) and one without (b$\to$c)
  electron exchange, become possible. 
  For FM 
  spins (triplet channel) or spinless fermions, however, 
  these two processes (a$\to$d) and (e$\to$h) cancel exactly because of the
  relative Fermi sign in the 
  next-nearest-neighbor hopping process. 
 \label{fig:dws_pm}}
\end{figure}

\section{Hubbard-Wigner Model}\label{sec:model}

The Hubbard-Wigner Hamiltonian investigated in this paper is motivated by the 
one-dimensional edge-sharing CuO-chains.~\cite{Horsch05,Mayr06} 
Edge-sharing chains are formed by CuO$_4$ squares just as in the CuO planes of 
the high-T$_c$ compounds, but these units are differently linked.
The edge-sharing arrangement leads to small nearest-neighbor hopping
matrix element  $t_1$
due to the almost 90$^\circ$ Cu-O-Cu coordination, and some contribution 
to $t_1$ stems from direct Cu $d$-$d$ overlap.~\cite{hopp}  Moreover, the
structure leads to a comparatively large matrix element $t_2$ between
second neighbor Cu ions stemming from a Cu-O-O-Cu
path.~\cite{Miz98} 
Thus edge-sharing chains, in contrast to the 180$^0$ bonded high-T$_c$ cuprates,
fulfill the fundamental criterion for
a WL, namely that the kinetic energy is small compared to the nearest-neighbor 
Coulomb interaction, in an optimal way.
While  the Coulomb repulsion is  screened by a static dielectric
constant in these insulators, the one over distance decay of the
interaction is preserved and 
must be taken into account. Truncation of the interaction may have serious
consequences for the charge-order pattern of generalized WLs \cite{Hubbard78}
as well as for their charge excitations.~\cite{Mayr06,daghofer:125116} 


The relevant states of Cu that need to be included in a low-energy model
are: Cu$^{3+}$
or, more precisely, the Cu $d^9$-ligand hole singlet state,   Cu$^{2+}$ with spin-1/2,
and  Cu$^{1+}$, corresponding to the filled $d$ shell. These
states can be expressed in the frame of a single-orbital Hubbard model with 0,1 or 2 electrons
per site.
Thus we consider the Hubbard-Wigner Hamiltonian, where the extension Wigner 
indicates that the long-range Coulomb interaction is included.
This model  has the form \cite{Horsch05,Mayr06}
\begin{equation}\begin{split}\label{eq:hamiltonian}
H & = - t_1 \sum_{i,\sigma} ({c}^\dagger_{i, \sigma}{c}^{\phantom{\dagger}}_{i+1, \sigma} + \textrm{h.c.})
    - t_2 \sum_{i,\sigma} ({c}^\dagger_{i, \sigma}{c}^{\phantom{\dagger}}_{i+2, \sigma}  + \textrm{h.c.}) \\
  &\quad +  U \sum_i n_{i, \uparrow}n_{i, \downarrow}
   + \sum_{l = 1}^{L/2} V_l \sum_i (n_i- \bar n)(n_{i+l}-\bar n)\;,
\end{split}\end{equation} 
where the 
operators $\cdag_{i,\sigma}$ ($\cnod_{i,\sigma}$) create (destroy)
electrons with spin $\sigma$ at lattice site $i$ with $i=1\dots L$.
The local density is given by $n_{i,\sigma} =
\cdag_{i,\sigma}\cnod_{i,\sigma}$, 
$n_{i} = n_{i,\uparrow} + n_{i,\downarrow}$, and the average density
is $\bar n=N_e/L$ for $N_e$ electrons.
The kinetic energy term includes 
NN hopping $t_1$ and NNN
hopping $t_2$, which are both typically much smaller than either the on-site
Coulomb repulsion $U$ or the NN-Coulomb interaction $V$ that parametrizes
the long-range Coulomb interaction $V_l = V/l$. In the case of finite rings
of length $L$, we
truncate the $1/l$ behavior at $L/2$, which is equivalent to
replacing $V/l$ by 
$\max(V/l,V/(L-l))$ for $0<l<L$. 
We have verified that small
modifications to the $1/l$ behavior do not affect our results. In
fact, truly long-range Coulomb repulsion is not crucial for the results
presented here: At quarter filling, both the FM kinetic exchange and
the weakened charge order can also be seen for on-site $U$ and NN Coulomb
repulsion $V_1$ only. 
In the following, we consider the model with long-range Coulomb interaction
and use the NN Coulomb repulsion $V_1 =V$ as unit of energy. 
Without loss of generality, $t_1$ is chosen to be positive. 

The Hamiltonian in Eq.(\ref{eq:hamiltonian})
contains two ingredients that have been shown to favor FM
correlations in Hubbard-like models: Strong on-site and
longer-range 
Coulomb repulsion~\cite{PhysRevB.57.10609} and,
perhaps more importantly, NNN
hopping.~\cite{PhysRevB.58.2635,Daul97,Pie96,Pen96}
%
%
In this paper we address the most transparent instance of 
the WL, namely quarter filling $\bar{n}=0.5$ for Hamiltonian
(\ref{eq:hamiltonian}).
We explore the magnetic properties
within the  WL regime, i.e., $t_1,|t_2| \ll V \ll U$,
which have not been explored before, and find FM ground states in a region of the
$t_1$-$t_2$ phase diagram with negative $t_2$.
Quite unexpectedly, we also find a strong influence of magnetism on WL
charge order. 
For comparison, we will first discuss the 
charge ordering for spinless fermions at half filling, corresponding
to the fully spin-polarized case with $\bar{n}=0.5$. At 
small $t_1$ and $t_2$, the alternating charge order is very rigid and its
lowest charge excitations are domain walls (DWs) with fractional
charge.~\cite{Jackiw75,Hubbard78,Rice82} DWs can be induced 
in a perfectly ordered state via NN hopping $t_1$, as
schematically illustrated in Fig.~\ref{fig:dws_pm}. 
Their creation
costs energy $\epsilon_0\sim V/2$ in the case of long-range Coulomb interaction,
and once created they can move easily through the
lattice via $t_1$ hopping processes. 
Their fractional charge  $\pm 1/2$ is responsible for 
the distinctive WL features in the optical conductivity 
and
in photoemission.~\cite{Mayr06,fratini:195103,daghofer:125116}

We investigate this model with exact diagonalization (ED) for spinless Fermions
and chains of up to $L=28$ sites. We use the Lanczos algorithm with a
numerical accuracy of $\approx 10^{-6}$ and check its validity by use
of full diagonalization for up to 18 sites. These ED calculations were
done for both the ground state ($T=0$) and for small but finite
temperatures ($T=10^{-4}-10^{-3}$), leading to practically identical results at
small $t_1$ and $t_2$
because there is a finite charge gap $\epsilon_0\approx V/2$, which blocks
changes at temperatures with an energy scale smaller than this.
For electrons with spin, we
use the DMRG with chains of
$L=24, 32,40$, and find 
results consistent with the ED for spinless fermions.  
In the DMRG, we keep 200 to 1200 states at each step, perform up to 10
finite-size sweeps and the neglected weight is $\lesssim
10^{-5}$. For parameters with a FM ground state, the energy obtained
with the DMRG agrees with the ED result.

\section{Charge order}\label{sec:charge}

\begin{figure}
  \includegraphics[width = 0.49\textwidth]{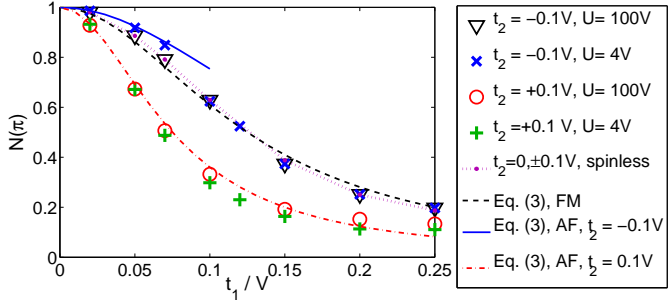}
  \caption{(Color online) Charge structure factor $N(q)$ for $q = \pi$
    as a function of nearest-neighbor 
    hopping $t_1$. The dotted line for spinless fermions is from ED
    calculations with $L=28$; 
    symbols were calculated using the DMRG with
    $L=24$ for electrons with spin.
    Analytic results are obtained from Eq.\ (\ref{eq:charge_cf2}) with $\Delta=\epsilon_0$ (FM)
    and $\Delta=\epsilon_0-2 t_2$ (AF), respectively.
    \label{fig:nn_AF_FM}}
\end{figure}

Increasing the NN hopping $t_1$ gradually reduces 
the charge ordering~\cite{Mayr06} until, at $t_1 \sim 0.2V$, the
charge gap vanishes.~\cite{Capponi00}
This is reflected in the charge structure factor 
\begin{equation}\label{eq:N_q}
N(q) = \langle \rho_{- q} \rho_{ q} \rangle \textnormal{, with\ } \rho_{ q} = 1/N_e\sum_r
\exp({-\textrm{i}  q  r}) n_r\; , 
\end{equation}
which, for perfect charge alternation, 
is peaked at $q = \pi$ with $N(\pi) = 1$.
As can be seen in Fig.~\ref{fig:nn_AF_FM}, the results for spinless
fermions, obtained using Lanczos diagonalization,
show that $N(\pi)$ is strongly reduced 
even before the gap vanishes, giving a weaker charge density wave.  We find
that, for spinless fermions,  the melting of WL charge order with $t_1$ does
\emph{not} depend on $t_2$. 
The behavior of 
$N(\pi)$ can be described
analytically because only  few 
DWs are present at small $t_1$. To leading
order, virtual DW excitations contribute  
\begin{equation}\label{eq:dw_energy}
E_c\sim - N_e 2 t_1^2/\epsilon_0 
\end{equation}
to the ground state energy.  
With $N_e = L/2$, we obtain 
\begin{equation}\label{eq:charge_cf2}
N(\pi) \simeq  \frac{1}{1+(4 t_1/\Delta)^2}\;,
\end{equation} 
given the charge gap $\Delta = \epsilon_0\sim V/2$. 
This expression is
indicated by the dashed line in Fig.~\ref{fig:nn_AF_FM} and agrees 
with the numerical data.

\begin{figure}
  \includegraphics[width = 0.47\textwidth]
  {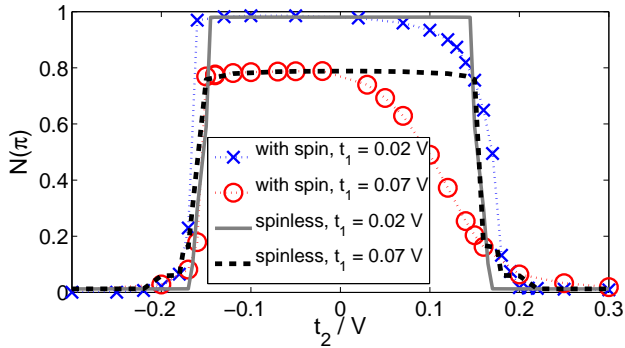}
  \caption{(Color online)  The charge structure factor $N(\pi)$  
    versus $t_2$.   Note that it does depend on the sign of $t_2$ 
    for fermions with spin and does not for spinless fermions. 
    The results for spinless fermions were calculated using exact diagonalization
    ($L=18$), and the results for electrons with spin using the DMRG ($t_1=0.02V$,
    $U=4V$, $L=24$ and $t_1=0.07V$, $U=100V$, $L=32$).
    \label{fig:nn_pi_pi2}} 
\end{figure}

\begin{figure}
  \includegraphics[width = 0.47\textwidth]{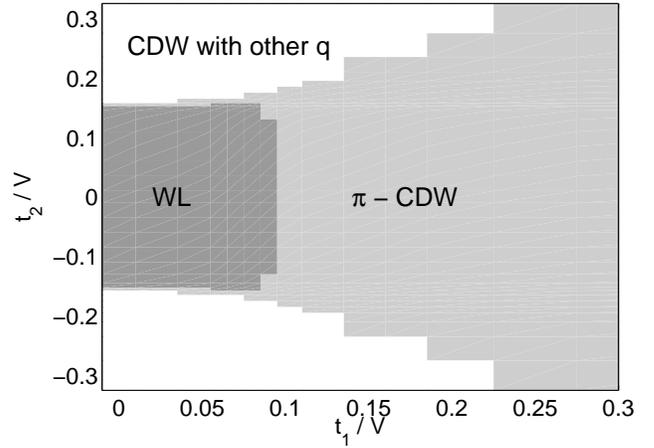}
  \caption{Phase diagram for spinless fermions determined from the charge
    structure factor $N(q)$, see Eq.\ (\ref{eq:N_q}). 
    WL (dark gray): strongly charge-ordered WL with $N(\pi) >
    0.7$. $\pi$-CDW (light gray): CDW with periodicity $\pi$, but
    $N(\pi) < 0.7$. (This choice corresponds approximately to the
    inflection point of $N(\pi)$ as a function of $t_1$.) In the white
    area, $N(q)$ has its maximum at $\pi/2 \leq q < \pi$, at $\pi/2$
    for large $t_2$. 
    In the exact diagonalization, we take $N_e=8$ fermions on $L=16$ sites. 
\label{fig:phases_nospin}}  
\end{figure}

In marked  
contrast to the gradual change that occurs with $t_1$, NNN hopping
$t_2$ is frustrated 
for spinless fermions until $N(\pi)$ drops sharply at a level-crossing
transition at $t_2^c \sim0.15V$, \cite{Mayr06} see also
Fig.~\ref{fig:nn_pi_pi2}. 
At  
the level crossing, the  ground state changes fundamentally; 
$N(q)$ develops a broad continuum with
a maximum between $\pi$ and $\pi/2$ (moving to
$\pi/2$ at large $t_2$) rather than at $\pi$.
Just as $t_2$ does not influence the charge order, weakening with $t_1$, the
level-crossing transition driven by $t_2$ is hardly affected by
$t_1$. This can be seen by comparing the $t_1=0.02 V$ and $t_1=0.07 V$
curves for spinless fermions in Fig.~\ref{fig:nn_pi_pi2}. 
Consequently, the WL phase is bounded by vertical and 
horizontal lines in the $t_1$--$t_2$ plane, see Fig.~\ref{fig:phases_nospin}. 

While the transition between the two CDW phases with $q=\pi$ and $q
\neq \pi$ 
depends on both $t_1$ and $t_2$, it is remarkable
that the WL is never affected by 
the \emph{combination} of hopping processes. 
We would 
actually expect some cooperative effects between $t_1$ and $t_2$ because NNN
hopping is no longer frustrated in the presence of $t_1$, see 
Fig.~\ref{fig:dws_pm}. Due to the DW delocalization,
(b$\leftrightarrow$c), two-DW states should gain energy
with $t_2$, and nonzero $t_2$ should thus help destabilize the
charge ordering.  The solution is found in the process shown in
(f$\leftrightarrow$g): For spinless fermions
(all arrows in Fig.~\ref{fig:dws_pm} pointing up), 
process (a$\leftrightarrow$e)
and process (f$\leftrightarrow$g) are equivalent. 
Since two electrons swap
places in the second case, the resulting Fermi sign leads to destructive
interference and the lowest-order processes associated both with
$t_1$ and with $t_2$ cancel out. 

After this discussion of the spinless model, we now turn to electrons
with spin. Due to 
the dominance of the Coulomb repulsion and the classical nature of WL
ordering, we might  
not expect charge ordering to be affected significantly
by the spin degree of freedom as long as $U \gg V$. 
However, the behavior of $N(\pi)$ obtained using the 
DMRG 
for electrons with spin indicates that there is a
surprisingly strong influence 
even for $U = 100V$. 
In contrast to spinless fermions, where $t_2$ does not affect the
behavior of $N(\pi)$ as a function of $t_1$, we find the 
charge order to be considerably weakened at $t_2 > 0$ for electrons with
spin, see Fig.~\ref{fig:nn_AF_FM}. 
We can understand this by considering the processes of Fig.~\ref{fig:dws_pm}:
The states depicted in (c) and (g) differ by their sequence of up and down spins. 
Process (b$\leftrightarrow$c) is then no longer
canceled by (f$\leftrightarrow$g), as it is for
spinless fermions. Consequently, a
kinetic energy contribution 
$\propto t_1^2t_2/\epsilon_0^2$  
is no longer forbidden by the Pauli principle.

Our interpretation is corroborated by analytic considerations: The
additional DW motion due to $t_2$ favors two-DW states and changes the 
gap relevant to Eq. (\ref{eq:charge_cf2}) from
$\Delta=\epsilon_0\sim V/2$ to $\Delta=\epsilon_0-2 t_2$. 
This leads to the
dash-dotted line in Fig.~\ref{fig:nn_AF_FM}, which indeed describes
the weakened charge order seen in the DMRG at $t_2 >0$. 
For $t_2 <0$, however, the DMRG results are described by the \emph{spinless}
gap $\Delta=\epsilon_0$. Since spinless fermions are equivalent to the fully
polarized FM state, this indicates \emph{ferromagnetism}, see below. 
For $U=4V$ and small 
$t_1 \lesssim 0.07 V$, 
where AF 
superexchange $\sim 4t_2^2/U$ destroys the polarized state, processes 
$\propto t_1^2t_2$ 
retain their impact and \emph{strengthen} charge order, see the full
line in Fig.~\ref{fig:nn_AF_FM}.

For fermions
with spin, the sharp transition as a function of 
$t_2$ shown in Fig.~\ref{fig:nn_pi_pi2} becomes asymmetric with
respect to the sign of $t_2$. 
Even for very small $t_1 = 0.02V$, the cooperation between $t_1$
and $t_2$ is enough to render the 
breakdown of the WL 
charge order
more gradual for $t_2 > 0$ than for $t_2 < 0$. For $t_1 = 0.07 V$, 
charge order is strongly reduced for $t_2 > 0$, and the sharp
drop in $N(\pi)$ as a function of $t_2$ has disappeared, in stark
contrast to the spinless model.

\section{Magnetism}\label{sec:spin}

\begin{figure}
  \includegraphics[width = 0.47\textwidth]{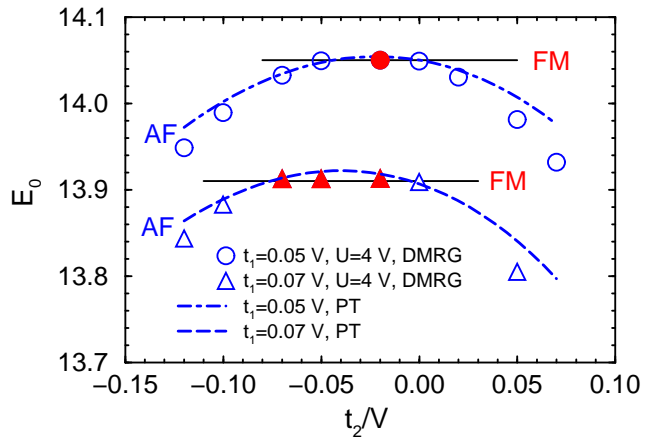}
  \caption{(Color online) Ground state energy $E_0$ for $L=24$ versus $t_2$ for $U = 4V$
    with $t_1 = 0.05V$ (circles) and $t_1 = 0.07V$ (triangles). Filled symbols
    indicate  a fully polarize ground state and horizontal lines the energy of
    the FM state at $t_2 = 0$. The dashed and dash-dotted lines are analytic
    results obtained from perturbation theory, Eq. (\ref{eq:magn_en}), see text.
    \label{fig:e_AF_FM}} 
\end{figure}

One expects AF interactions for the WL~\cite{Valenzuela03}  due to
superexchange in the generalized Hubbard model. Yet our observation
that the charge structure factor at 
negative $t_2$ agrees closely with results for spinless 
fermions, see Figs.~\ref{fig:nn_AF_FM} and~\ref{fig:nn_pi_pi2}, is
already an indication that the ground state in this regime is FM.
This is indeed the case, and our DMRG studies in fact yield fully
polarized ground states for some parameter sets at negative $t_2$. As
can be seen in Fig.~\ref{fig:e_AF_FM}, the FM interval increases with $t_1$.

In the following,
we  analyze the magnetic exchange by using perturbation theory valid
when $t_1,t_2 \ll V \ll U$. The robust WL charge order leads 
to a \emph{modulated} Heisenberg chain with spins at every second
lattice site. Magnetism is therefore described by an effective
Heisenberg-like Hamiltonian   
\begin{equation}\label{eq:heis}
H_J  = J \sum_{i} ({\bf S}_i\cdot{\bf S}_{i+2}-\frac{1}{4}n_i n_{i+2})\;,
\end{equation} 
where $i$ runs only over the even sites, where the $L/2$ spins forming
the WL are located.
The total ground-state energy then is $E=\langle H_J\rangle +E_{FM}$,
where $E_{FM}$ is the energy of the fully spin-polarized state, which
is
equivalent to the ground-state energy of spinless fermions.
There are two distinct mechanisms that contribute to the exchange constant
$J=J_{SE}+J_{KE}$. The first term is the usual superexchange, which
involves a doubly occupied intermediate state and therefore has 
the energy scale
%
$U$ in the denominator: 
\begin{equation}\label{eq:J_SE}
J_{SE}\simeq \frac{4 t_2^2}{U}+ \frac{12 t_1^4}{\epsilon_0^2 U}+
\frac{8 t_1^2 t_2}{\epsilon_0 U}+ \dots
\end{equation} 
The second term 
$J_{KE}$---denoted as 
the {\it kinetic exchange}---arises from a
spin exchange without any doubly occupied sites, i.e., 
exactly from the same effect that weakens the charge order for $t_2 >0$: 
Quantum interference between processes (a$\leftrightarrow$e) and
(f$\leftrightarrow$g) in Fig.~\ref{fig:dws_pm} is \emph{destructive} in the
polarized FM state and \emph{constructive} in the AF singlet, which 
leads to an exchange energy
\begin{equation}\begin{split}\label{eq:J_kin}
J_{KE} \simeq \frac{2t_1^2}{\epsilon_0}\left(\frac{1}{1-{2t_2}/{\epsilon_0}}-1\right)
\simeq \frac{4 t_1^2 t_2}{\epsilon_0^2}+\dots 
\end{split}\end{equation} 
that depends on the sign of $t_2$.
The NN correlation function of the 1D quantum
antiferromagnet is given by $\kappa = - \ln 2 + 1/4 \approx -0.443$,
corresponding to $\langle {\bf S}_i\cdot{\bf S}_{i+2} \rangle$ for our
modulated chain. Inserting $\kappa$ as well as Eqs.\ (\ref{eq:J_SE}) and\ (\ref{eq:J_kin}) into
Eq.~(\ref{eq:heis}), we obtain an analytic estimate, valid at small
hopping, for the energy of the AF state 
\begin{equation}\label{eq:magn_en}
E_{AF}\simeq \frac{L}{2} (J_{SE}+J_{KE}) (\kappa-1/4)+E_{FM}\;.
\end{equation}
We compare this analytic result to 
numerical DMRG data for $t_1=0.05 V$ and $t_1=0.07 V$, 
in Fig.~\ref{fig:e_AF_FM}. This figure shows 
%
that the analytic curves
given by $E_0=\min(E_{AF},E_{FM})$ closely model the numerical ground-state
energies for not-too-large $t_2$ with no fitting parameters.
Moreover, the analytical boundaries of the
FM phase, namely $t_2^a \sim -3 t_1^2/U $ and $t_2^b \sim -(U/\epsilon_0^2)
t_1^2 $, match the magnetic phase boundaries found using the 
DMRG.

This may be compared with the boundaries in the $t_1$-$t_2$ plane
describing the appearance of four Fermi points,
namely $t_2 \leq  -(t_1/4)\sec(\pi n/4)^2$ and  
$t_2 \geq (t_1/4)\csc(\pi n/4)^2$, where $n$ is the filling. For
the quarter-filled case, i.e., $n=1/2$, only the first relation is of interest
and yields the condition $-\infty\leq t_2\leq -0.293 t_1$ for the appearance
of FM in the Hubbard model, as observed in Refs.\ \onlinecite{Pie96}
and \onlinecite{PhysRevB.58.2635}.
It is perhaps not surprising that the boundary relevant 
for the Hubbard model, which is linear in $t_1$, is completely different from the boundaries
obtained for the Wigner lattice, which are both quadratic in $t_1$.
Nevertheless, ferromagnetism at quarter-filling
is found at negative $t_2$ in both cases!

\begin{figure}
  \includegraphics[width = 0.49\textwidth]{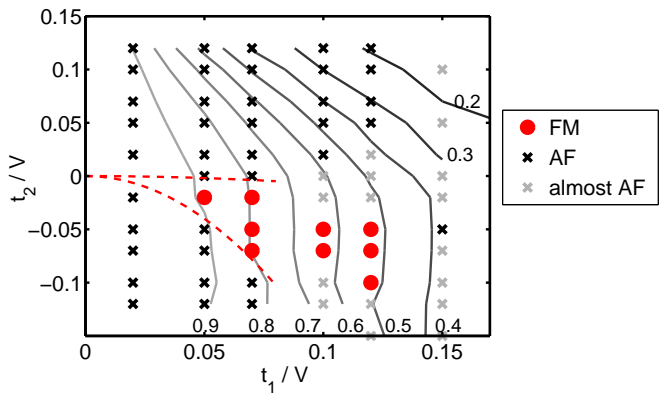}
  \caption{
    (Color online) Magnetic phases and charge structure in the WL
    regime.  
    FM ($\circ$): $m > 0.99M$; AF (black $\times$): $m <
    0.01M$; `almost AF' (gray $\times$): $0.01M\leq m < 0.1M$, where
    $m = S_\textrm{tot}(S_\textrm{tot}+1)$ was obtained by DMRG for
    $L=24, U=4V$, and $M = S_\textrm{max}(S_\textrm{max}+1)=42$.
    Gray lines give potential lines for $N(\pi)$, i.e., the strength of the
    alternating charge order. The lines were interpolated from data points
    obtained at the same parameter values as the magnetic data.\label{fig:phases_mag}} 
\end{figure}

The phase diagram in Fig.~\ref{fig:phases_mag} shows the total spin in
the ground state of a chain 
with $L=24$ for the range of $t_1, t_2$ corresponding to the WL regime. 
The dashed lines represent the analytical boundaries $t_2^a$ and $t_2^b$
of the FM region.
Kinetic exchange (\ref{eq:J_kin})---the only magnetic interaction surviving
for $U\rightarrow \infty$---is  AF for $t_2>0$. 
%
For $t_2 < 0$, $J_{KE}$ raises the 
energy of the AF state
over that of the FM state, see Fig~\ref{fig:e_AF_FM}. 
In the FM state itself, this effective FM
exchange is, ironically, \emph{absent}, because the Pauli principle forbids the
ring exchange in the polarized state. At large negative $t_2$ and 
not-too-large $U$, AF superexchange (\ref{eq:J_SE}) once again
dominates. 

The phase diagram Fig.~\ref{fig:phases_mag} also contains contour lines
$N(q=\pi, t_1, t_2) =C $ with the constant $C= 0.9, 0.8, ..., 0.4$,
which indicate the strength of the alternating ($q = \pi$) charge order
of the Wigner lattice.  
In accordance with Figs.~\ref{fig:nn_AF_FM} and~\ref{fig:nn_pi_pi2}, we find
that the charge order dies off 
most quickly in the singlet states in the $t_2 > 0$ region. 
The analytic contour lines 
$t_1^c$ versus $t_2$ 
following from Eq.\ (\ref{eq:charge_cf2}) for $t_2>0$  have 
%
the form 
\begin{equation}\label{eq:contour}
t_1^c \simeq \frac{1}{4} (\epsilon_0-2 t_2) \sqrt{1/N(\pi)-1}\;.
\end{equation}
They agree with the DMRG data,
just as Eq.\ (\ref{eq:charge_cf2}) agrees with 
$N(\pi)$ in Fig.~\ref{fig:nn_AF_FM}. 
Since our analytic result
describes the unbiased DMRG simulations so well, we conclude that the kinetic
exchange indeed drives the suppression of charge order in the AF regime for
$t_2 > 0$. 

\section{Discussion and Conclusions}\label{sec:conclusions}

We have investigated charge order and magnetism of the 1D
quarter-filled Wigner lattice with nearest and next-nearest neighbor hopping. 
Starting from the regime $t_1,|t_2| \ll V$ with  extremely strong alternating charge order
stabilized by the long-range Coulomb 
repulsion, we find that increasing NN hopping $t_1$ drives a crossover
to a $4k_F$ charge density state with weaker charge order but unchanged modulation
period, 
whereas increasing NNN hopping $t_2$ leads to a
sudden level-crossing transition, destroying 
the alternating charge
order.\cite{Mayr06} For $t_1,|t_2| \ll V$ and spinless fermions, we find that there are
no mixed processes involving both $t_1$ and $t_2$ because destructive
interference removes the lowest order processes $\propto t_1^2
t_2^{\phantom{2}}$. 
Consequently, the WL is bounded by a vertical crossover line and 
horizontal phase transition lines in the 
$t_1$-$t_2$ phase diagram for spinless fermions, see
Fig.~\ref{fig:phases_nospin}.  

However, in the case of real electrons with spin, 
we find that processes $\propto t_1^2
t_2^{\phantom{2}}$ are absent in the FM state
(as for spinless fermions) but not in
the AF state (as illustrated  in Fig.~\ref{fig:dws_pm}).
This results in an effective magnetic exchange
$\simeq 4 t_1^2 t_2/\epsilon_0^2$
which favors the AF relative to the FM state for positive $t_2$
and disfavors the AF state for negative $t_2$. 
This peculiar effect  is corroborated by DMRG data, where we indeed
find AF as well as FM ground states which depend on $t_1$ and $t_2$.
The phase boundaries of the FM phase obtained here for the quarter-filled
WL are distinct from those obtained for the $t_1$-$t_2$ Hubbard models 
driven by large $U$ and Fermi surface topology, where ferromagnetism is
found whenever the fully polarized Fermi sea is split in
two.~\cite{Pie96} 
%
%
This is perhaps not that surprising, as the
charge order in the WL is not caused by a quantum mechanical
Fermi-surface instability but by the strong  Coulomb
repulsion, which does not depend on the Fermi surface. 
Thus, the situation in the WL is very different from charge and
magnetic order driven by Fermi surface instabilities in the Hubbard model
with purely on-site Coulomb repulsion, which was studied extensively in
Ref.~\onlinecite{schuster:045124}.

Magnetism in the WL  is not driven by Fermi-surface instabilities, but
actually has more in 
common with the FM reported for a model of coupled chains with a symmetry-breaking on-site
potential.~\cite{Pen96} In these coupled chains, charge
order is stabilized by a strong on-site potential, which corresponds
to the spontaneous symmetry breaking by long-range Coulomb repulsion
in the case of the WL, and is likewise independent of the Fermi
surface. Starting from strong charge order, we have been able to derive the
effective magnetic exchange terms using perturbation theory, and
find the resulting magnetic energy to be in good agreement with DMRG data,
see Fig.~\ref{fig:e_AF_FM}. The most important terms in the effective
exchange Hamiltonian
are the AF superexchange $\propto t_2^2/U$ involving a doubly
occupied site and the \emph{kinetic} exchange term $\propto
t_1^2t^{\phantom{2}}_2/V^2$. A related FM exchange 
mechanism based on strong charge ordering and NN hopping has been 
invoked for two-dimensional kagome lattices.~\cite{Pollmann07}

Since neither charge order nor magnetism are driven by a Fermi surface
instability but are determined by the Coulomb interactions,  
one might not expect that the charge order depends significantly
on the magnetic correlations.
Indeed, Wigner lattices are
often discussed in terms of spinless fermions, and the magnetic
exchange is added only in terms of a modulated Heisenberg model for the
given charge order. 
Our calculations show 
that this picture is too simple: Figures~\ref{fig:nn_AF_FM}
and~\ref{fig:nn_pi_pi2} reveal that magnetic correlations have a very
strong impact indeed on charge order. In fact, the weakening of
charge order in the AF phase at positive $t_2$
is due to
the same processes $\propto t_1^2t^{\phantom{2}}_2$ that cause the
kinetic exchange. 
While these virtual processes cancel in the FM phase,
they involve domain-wall excitations in the AF phase, see the sketch 
in Fig.~\ref{fig:dws_pm}. Controlled by the $t_2$ dependent two-DW gap $\Delta$,
the admixture of virtual DW excitations leads to a reduction of
charge order in the AF state at positive $t_2$, as can be
seen in the phase diagram Fig.~\ref{fig:phases_mag}.
Within the FM regime at negative $t_2$, charge order does not
depend on $t_2$. At large negative $t_2$, however,
the antiferromagnetic phase
reappears, but now, surprisingly, the charge order becomes stiffer
as $|t_2|$ is increased. This peculiar behavior finds its explanation 
in the $t_2$ dependence of the domain-wall gap $\Delta$ in the AF state.

In summary, we have shown that, in Wigner lattices with next-nearest-neighbor
hopping $t_2$, a \emph{kinetic exchange mechanism} 
is at work, which favors
ferromagnetism for negative NNN hopping $t_2$ and  might explain the
spin polarization recently observed in strongly 
charge-ordered carbon nanotubes.~\cite{Des08} 
In contrast to the usual superexchange processes, which involve virtual
excitations across the Mott-Hubbard gap $\sim U$, kinetic exchange arises
from virtual transitions across the Wigner lattice charge gap 
$\epsilon_0\propto V \ll U$.
Although Fermi surface
topology describes neither charge nor magnetic order in the strongly
correlated WL, we 
nevertheless
find that quantum interference of 
electrons is important in the WL. In fact, the spin degrees of freedom
have such a strong impact in the AF regime
that charge ordering \emph{cannot} be described reliably in
terms of spinless fermions, even  in the extreme WL regime $t_1 , |t_2| \ll V \ll U$. 
As this effect is intimately related to the \emph{kinetic exchange mechanism},
it may also be relevant in higher dimensions, e.g., the
above-mentioned kagome systems.

\acknowledgments
We thank D. Baeriswyl, K. Hallberg,
M. Jansen, N. Kawakami, B. Keimer, G. Khaliullin, 
S. Maekawa, W. Metzner, C. Penc, R. Zeyher and T. Tohyama for useful
discussions.
This research (M.D.) was partly supported by the NSF under grant DMR-0706020.


\begin{thebibliography}{10}
\bibitem{hopp}
A typical size of $t_1$ is 50 meV in edge sharing chains as compared to 500 meV
  in 180 degree bonds as in the high-$T_c$ or ladder compounds.
\end{thebibliography}

\end{document}